**Persistent pods of the tree *Acacia caven*: a natural refuge for diverse insects including Bruchid beetles and the parasitoids Trichogrammatidae, Pteromalidae and Eulophidae.**

D. Rojas-Rousse

*Institut de Recherche sur la Biologie de l'Insecte, UMR CNRS 6035, Faculté des Sciences, F-37200 Tours, France*
rousse@univ-tours.fr

**Abstract**

The persistent pods of the Leguminosae *Acacia caven* (Mol.) that do not fall from the tree provide opportunities for the appearance of a diverse group of insects the following season. Such pods collected during the spring of 1999 in Chile were indehiscent with highly sclerified pod walls. In contrast, persistent pods collected in Uruguay after a wet winter and spring (2002) were partially dehiscent, inducing the deterioration of the woody pods, and consequently exposing the seeds. These persistent pods are a natural refuge for insect species, namely two bruchid beetles (*Pseudopachymeria spinipes, Stator furcatus*), one scolytidae (*Dendroctonus* sp), lepidopterous larvae, ant colonies (*Camponotus* sp), one species of oophagous parasitoid (*Uscana espinae* goup *senex*), the gregarious larval-pupae parasitoid *Monoksa dorsiplana* (Pteromalidae) and two species of *Horismenus spp.* (Eulophidae). The patriline of *M. dorsiplana* is frequently formed by 1 son + 7 daughters.

**Introduction**

*Acacia caven* (Mol.), a Leguminosae belonging to the sub-family Mimosoideae, is a tree species of extra-tropical South America, found between 36°N and 18°S and from the Atlantic to the Pacific. It is usually found in the continental climate of the Grand Chaco region of northern Argentina, central Paraguay and southern Bolivia, as well as in the Mediterranean-type climate zone of Chile, a few parts of southern Brazil and part of Uruguay, where it is often thoroughly integrated with various species of subgenera of *Acacia, Prosopis* and *Schinopsis* (Aronson and Ovalle, 1989). *A.caven* is colloquially known as "espino" or

"espinillo", and is a prominent woody invader in over-grazed cow pastures or abandoned fields. It seems that the espinales are a recent phenomenon in the central valley of Chile, i.e. no more than 500 – 2000 years old, probably introduced from northern Argentina by guanacos, pack animals (llamas) in Trans-Andean caravans (Ovalle *et al.*, 1990, Muñoz *et al.*, 1993). *A. caven* is one of the most specialized members of the legume subfamily Mimosoidea, by virtue of its large colporate pollen, diploid chromosomes, spinescent stipules, and multiseriate, indehiscent pod construction. It is related to, and often confused with, the pantropical *Acacia farnesia* (L.) Willd., which is not found in Chile (Ovalle *et al.*,1990). Its phenological behavior is distinctly out of phase with that of all woody plants native to this Mediterranean-climate area. The 'reversed' phenological pattern exhibited by *A. caven* in Chile was probably a relictual development, which evolved in harmony with the climatic conditions prevailing in the subtropical, dry winter climate of the Chaco, prior to its migration to central Chile in Quaternary times (Aranson *et al.*, 1994). *A. caven* loses its leaves at the beginning of the rainy season and puts out new leaves at the beginning of the dry season. It flowers in the spring before the appearance of the leaves, which develop completely during the summer. The pods ripen in autumn (Muñoz *et al.*, 1993). The majority of indehiscent pods fall to the ground, but some remain on the trees for several months.

Larvae of the bruchid *Pseudopachymeria spinipes* feed mostly on the seeds of *Acacia* (Miller). Although *P. spinipes* has been imported "into various parts of the Old World, its apparent original distribution was in Argentina, Brazil, Peru and Ecuador" (Johnson and Siemens, 1997). It is also found in Chile where its only host is *A. caven* (Barriga Tuñon, 1990).

In Chile, analysis of phenological relations between the cycle of *A. caven* and that of *P. spinipes* shows that one generation of *P. spinipes* develops within 10-11 months. Adult longevity is about 20 days, and females oviposit on developing pods, eggs being laid aggregated (Saiz *et al.*, 1977; Avendano and Saiz, 1978; Francisco Saiz *et al.*, 1980). The new larvae enter the developing pods thanks to their very hard chitinized mouthparts (Avendano and Saiz, 1978; Francisco Saiz *et al.*, 1980). The newly hatched adults mate immediately after emergence (sometimes even before they leave the pod), then disperse to search for appropriate pods for oviposition. However, as egg-laying behavior is inhibited by the presence of flowers on the host plant and is triggered by the presence of pods (Yates and Saïz, 1989), the females oviposit on the persistent pods of the previous season which have not fallen. These pods allow a high level of bruchids to appear during the following season (Saiz, 1993).

Currently, the only known natural enemies of *P. spinipes* are a microhymenopterous parasitoid mentioned in a study on agronomical bruchidae pests in Chile (Barriga Tuñon, 1990), and an oophagous parasitoid from the genus *Uscana* collected once near Santiago (1981) and once in 1995 near Cauquenes (VII region), identified as *Uscana espinae* (Pintureau and Gerding) group *senex* (Pintureau *et al*., 1999).

As persistent pods of *A.caven* provide a refuge for the population of *P. spinipes* (Saiz, 1993), we investigated the presence of this species in the persistent pods, as well as the presence of other insects liable to compete with or to parasite *P. spinipes*. Pods were collected in Chile in November 1997 (region VIII) and in November 2002 in Uruguay (Prado Parc, Montevideo).

**Material and Methods**

*Biodiversity of insects living in persistent pods of A.caven.*

Ripe pods of *A. caven* persisting on trees were collected during the spring (November) of 1997 (Chile: region VIII) and 2002 (Uruguay: Prado Parc, Montevideo). Each pod was labeled with its place of origin and date of collection, measured (length in cm), and placed in a box covered with a thin cloth. Boxes were stored in a climatic chamber with a 12h / 12h photoperiod and a 30°/ 20°C thermoperiod.

Holes pierced by bruchid beetles were observed on the pod integument, and adult bruchid beetles and parasitoids were counted during a two-month period. Thereafter, each pod was opened to count the number of seeds, and the seeds were kept to count any remaining emergent insects.

*Dynamic Population of Bruchidae and parasitoids developed in one persistent pod*

As only one *P. spinipes* adult can develop in a single seed, the newly emerged population of bruchid beetles corresponds to the number of holes in the seeds (Saiz *et al*., 1977). If only the holes pierced on the pods are counted, the size of the population will be underestimated (Saiz *et al*., 1977). The difference between the total number of seeds in each pod and the number of seeds attacked corresponds to the quantity of seeds able to start growth and allow the spread of *Acacia caven.*

No parasitoid emergence holes were observed on the pods. However, after the rainy season, some of the pods were partially dehiscent, allowing parasitoid adults to leave the pod easily. They could also have passed through the holes pierced on the pod's integument by the bruchid beetles. As only one *P. spinipes* adult can develop from a single seed, the level of parasitism is apparent from the number of parasitoid emergence holes observed on the seeds.

*Gregarious behavior of Monoksa dorsiplana (Boucek)* (Pteromalidae)

The Pteromalidae *Monoksa dorsiplana (Boucek)* was first described in 1990 in pods of *Acacia farnesiana* collected in Israël (Boucek, 1990). In order to study some of its life history traits, its development on a substitution host was carried out successfully in the laboratory, whereas this transfer was not successful for the Eulophidae *Horismenus spp.*. The mass rearing of *M. dorsiplana* was carried out on *Callosobruchus maculatus* (Bruchidae) reared on *Vigna radiata* (Leguminosae). The hosts and parasitoids were mass-reared under conditions close to those of their zone of origin: 30°/ 20°C, 12:12-h L:D, and 70% RH, with synchronous photo- and thermo-periods.

The offspring of 63 inseminated *M. dorsiplana* females were studied. Newly emerged females were individually placed with 2 males in Petri dishes for 24h to ensure mating. Females were then individually provided with one *V. radiata* seed containing either an L4 larva or pupa of *C. maculatus* (Huignard *et al.*, 1985). These females were given a new host every day until they were 5 days old. Every 24 hours, the seed offered to the female was removed and placed in an Eppendorf tube closed with a cotton cap until the birth of the offspring.

**Results**

*Biodiversity of the entomofauna associated with the persistent pods*

Only *Pseudopachymeria spinipes* (Erickson) emerged from pods collected in Chile, while the presence of *Stator furcatus* (Johnson) was also observed in those from Uruguay (identification by Clarence Dan Johnson of Northern Arizona University, USA). Some of the 43 pods collected in Uruguay after the rainy season contained other insect families. The pod apertures facilitated observation of the presence of adults and larvae of the scolyte *Dendroctonus* sp in the woody walls of the pods, and also lepidopterous larvae feeding on the seeds. In three pods, small nests of *Camponotus* sp. (Formicidae) were found (identified by Jean Luc Mercier, LEPCO,Tours, France).

Oophagous and larvae-pupae parasitoids were also collected. *P. spinipes* eggs were parasitized by *Uscana espinae* group *senex* (Pintureau and Gering) (Pintureau *et al.*, 1999), and the larvae were parasitized by *Monoksa dorsiplana* (Boucek) (Pteromalidae: identified by J. Yves Rasplus, INRA and Gérard Delvare, CIRAD, Montpellier, France) and by two *Horismenus* spp. (Eulophidae: identified by John Lasalle, Natural History Museum of London and Gérard Delvare, CIRAD, Montpellier, France).

*Pod size and quantity of seeds in a persistent pod.*

The pod length (cm) distribution was bimodal according to origin (Chile or Uruguay) (Fig. 1). The length of pods collected in Uruguay varied from 1 to 5.5cm (modal class: 2 – 2.5cm) (N=43: mean length ± standard error of the mean: 2.7± 0.3 cm), while those collected in Chile ranged from 3 to 8.5cm (modal class: 5 - 5.5cm) (N = 60: mean length ± standard error of the mean 5.4 ± 0.3 cm) (Fig. 1), i.e. twice as long as the Uruguayan pods (ANOVA: XLStats.6 for Windows: $F_{101,102}$ = 167.04, P < 0.0001). The linear regression between pod length and the quantity of seeds per pod was positive: only a couple of values (out of 103) were outside the second confidence interval (Simple linear regression XLStats.6 for Windows: R= 0. 62) (Fig. 2).

*Intensity of attack by the Bruchidae Pseudopachymeria spinipes and Stator furcatus*
   *Pods collected in Chile*

Around 68% of indehiscent pods (41/60) were attacked by *P. spinipes*, as shown by the presence of 1 to 5 emergence holes (Table 1).

**Table 1.** Intensity of attack by *P. spinipes* on persistent pods of *A.caven*.

|  | **Number of *P. spinipes* emergence holes per pod** | | | | | |
|---|---|---|---|---|---|---|
|  | 0 | 1 | 2 | 3 | 4 | 5 |
| **Pods observed (N = 60)** | 19 | 27 | 8 | 4 | 1 | 1 |
| **Total emerged bruchids** | 0 | 168 | 87 | 33 | 30 | 5 |

On the majority of attacked pods, we found only one *P. spinipes* emergence hole (27/41= 66%), and only 2.5% pods had 4 or 5 holes (1/41= 2.5%) (Table1). As only one *P. spinipes* adult can develop in a seed and the number of attacked seeds was greater than the number of emergence holes in the pods, we can infer that the majority of bruchids left the pod by the first hole(s) pierced (Table 1).

*Pods collected in Uruguay*

The bruchid beetles in pods collected after the rainy season at the end of November 2002 might have been able to leave the partially dehiscent pods without having to pierce the integument. The bruchid beetle attack was therefore analyzed by counting the emergence holes on the seeds of 50 pods.
Some pods (24%: 12/50) were contaminated by a second bruchid species: *Stator furcatus,* which is smaller than *P. spinipes* and lays 1 to 5 eggs on the integument (Fig. 3). In contrast to *P. spinipes,* 1 to 3 *S. furcatus* emergence holes were observed, indicating that more than one *S. furcatus* adult could develop per seed (Fig. 4). *P. spinipes* egg-laying was also observed in the split of partially dehiscent pods, the eggs aggregated on the pod integument. Thus, the total level of bruchid attack was the sum of the emergence holes of the two bruchids (*S. furcatus* + *P. spinipes*) and those of parasitoids (smaller than those of bruchids). The presence of parasitoid emergence holes indicated the presence of a bruchid host which had been parasitized.

Study of the change in seed quality indicated that the number of contaminated seeds was close to the total number of seeds (Fig. 5), with 82% of the total seeds (587 / 717) attacked by the bruchid beetles. On average, there were $14.35 \pm 1.45$ seeds in a pod (mean ± standard error), with $11.75 \pm 1.40$ contaminated by bruchid beetles (the difference observed was significant: Student test, $t = 6.10$; alpha $= 0.05$). The difference between the total number of seeds and the number of attacked seeds per pod corresponded to the seeds able to start growth (10.5%), and those which could not (7.5%).

*Parasitism of bruchids*

The percentage of parasitized seeds per pod represented the ratio of parasitized seeds / total attacked seeds (Table 2). The parasitism of bruchid beetles differed between pods ($\chi^2$ calculated $= 19.11$: alpha $= 0.01$ $\chi^2_{ddl\ 5} = 15.08$). In fact, in 17 pods attacked by bruchid beetles there was no parasitism, while 2 out of 50 attacked pods were parasitized from 80 to 100% (Table 2).

**Table 2.** Pattern of parasitism in 50 Uruguayan pods attacked by *Pseudopachymeria spinipes.*

|  | **Percentages of parasitized seeds in a pod (N = 50 pods)** | | | | | |
| --- | --- | --- | --- | --- | --- | --- |
|  | 0% | 1 to 20% | 21 to 40% | 41 to 60% | 61 to 80% | 81 to 100% |
| **Number of Pods** | 17 | 11 | 11 | 5 | 4 | 2 |

From all the pods attacked (Chilean + Uruguayan), 149 *Monoksa dorsiplana* females and 55 males emerged (proportion of females: 0.73) (Pteromalidae), and 384 females and 222 males of *Horismenus spp.* (Eulophidae) (proportion of females: 0.63). These parasitoids (total = 810) emerged from 285 seeds (with 1 *P. spinipes* host per seed), which could only be possible if the parasitoids had a gregarious larval development.

*Some life history traits of Monoksa dorsiplana*

The substitution host (*C. maculatus*) was fully adopted by 70% of egg-laying females (44 / 63), allowing total development of *M. dorsiplana*. The distribution of male and female offspring showed that one egg-laying female (provided with one host per day for 4 consecutive days) could have a maximum of 34 daughters for 9 sons (Fig. 6). The parasitoids emerging from one host allowed the most common associations to be studied (0, 1, 2 sons associated with 0, 1, 2 daughters etc.). The most common patriline of the offspring of one egg-laying female (provided with one host per day) was 1 male and 7 females (Figs. 7 and 8).

**Discussion**

The bruchid beetles attacking legume seeds, and the level of contamination, correlate to three patterns of pod dehiscence, classified as dehiscent, partially dehiscent and indehiscent (Johnson, 1981). According the pattern of dehiscence, the pods are attacked by three different guilds of bruchids, but this association of 'one pod dehiscence pattern/one species of bruchid beetle' is related to the different stages of pod or seed maturity (Johnson, 1981). In fact, all three bruchid guilds may attack one species of a partially dehiscent legume at different stages of pod or seed maturity, and indehiscent pods are sometimes tardily dehiscent after several months, the deterioration of the woody pods exposing the seeds (Johnson, 1981).

In Chile (region of Cauquenes), after a dry winter and spring (1997), all the persistent pods collected were indehiscent with highly sclerified walls. On the other hand, after a wet winter and spring, some persistent pods collected in Uruguay (2002) were partially dehiscent, inducing the deterioration of the woody pods, and consequently exposing the seeds. Such damaged pods allowed a diverse entomofauna to develop and/or to take refuge inside them. Hence, persistent pods of *A caven* provided a refuge for a variety of insect species, including two bruchid beetles (*P. spinipes* and *S. furcatus*), one scolytidae (*Dendroctonus* sp), lepidopterous larvae, ant colonies (*Camponotus* sp) and hymenopterous parasitoids (one oophagous parasitoid, *Uscana espinae* group *senex*, and larval-pupae parasitoids: *M. dorsiplana, Horismenus spp.*).

Of the indehiscent pods collected in Chile, with or without emergence holes, 68% were attacked by *P. spinipes*. However, the presence of emergence holes is not a reliable index of the intensity of bruchid beetle attack (Saiz *et al.,* 1990). In fact, a maximum of 1 to 5 emergence holes were counted per pod, whereas 168 *P. spinipes* adults emerged from 27 pods with only 1 emergence hole. This observation indicates that the majority of bruchid beetles left the pod by the first pierced hole. As the construction of pods is multiseriate, each seed is enclosed in a lodge, so in order to leave the pod, the bruchid beetles must drill the walls of several lodges to reach the first pierced hole. This behavior could be explained by a positive photo-attractivity induced by light going through the first pierced hole.

In Uruguay, the semi-dehiscent pods already attacked by *P. spinipes* showed that 24% were infiltrated by a second species of bruchid beetle, *S. furcatus,* presumably due to the deterioration of their woody walls. Around 82% of the seeds of pods attacked by *P. spinipes* + *S. furcatus* were destroyed. In contrast to *P. spinipes,* more than one *S. furcatus* adult can develop per seed, and we observed up to 3 adults.

The seeds of persistent pods of *A. caven* (which remained on the trees) were accessible to a guild of bruchid beetles after unfavorable climatic conditions which deteriorated the woody walls of the pods. The attack on seeds was most direct under these conditions, because they did not need to wait to lay their eggs for the transit of indehiscent pods (fallen on the substrate) through the digestive tract of native or introduced mammals (Johnson 1981, Traverset 1990). For example, in Central America the indehiscent pods of *A. farnesia* are eaten by horses, deer and ctenosaur lizards, and the seeds which pass intact in the feces are attacked by *S. vachelliae* (Traverset 1990).

These two species, *P. spinipes* and *S. furcatus*, constitute a significant stock of adults able to infect legumes. The *P. spinipes* adults emerging from persistent pods lay their eggs from January to March on the new green pods which the neonatal larvae, with hard chitinized mouthparts, are able to penetrate more easily than sclerified pods (Saiz et al., 1980).

Two native Trichogrammatidae were recently found as oophagous parasitoids of bruchid beetle eggs: *Uscana chiliensis* (Pintureau and Gering) on *Bruchus pisorum*, and *Uscana espinae* (Pintureau and Gering) on *P. spinipes* (Pintureau et al., 1999). In addition, one Pteromalidae (*M. dorsiplana*) and one Eulophidae (two species of *Horismenus spp.*) emerged from seeds of the persistent pods of A.caven attacked by *P. spinipes*. These persistent pods appear to be a refuge for bruchid parasitoids. This new information should facilitate the study of native parasitoids of South America for biological control of Bruchidae.

The species *Horismenus spp.* represents 59.1% of the parasitoid community of Bruchidae *Acanthoscelides* sp. and *Zabrotes sp*. (Hansson *et al.,* 2004). The genus *Horismenus* is predominantly a New World group, with its main distribution in the neo-tropical region. Currently there are 53 known species in the Americas and one species in Europe. The species are parasitoids or hyperparsitoids of Coleoptera, Diptera and Lepidoptera larvae. The majority of the species have so far not been described, and the identities of many of the 50 or so species which have been described are unclear (Hansson *et al.,* 2004).

The Pteromalidae *M. dorsiplana* was first described in 1990 (Boucek, 1990). This is the first time that this parasitoid has been found in South America on seeds of partially dehiscent persistent pods of *A.caven* attacked by *P. spinipes*. As only one bruchid beetle can develop in a seed, we conclude that this species is a gregarious larval ectoparasitoïd (proportion of females: 0.73). In contrast to *Horismenus spp., M. dorsiplana* was reared on a substitution host (*C. maculatus*) without any specific selection, since 70% of females produced offspring. The patriline of *M. dorsiplana* was generally composed of 1 son + 7 daughters.

The persistent indehiscent pods of *A. caven* which become partially-dehiscent under unfavorable climatic conditions (rainy season during winter and spring), provide a refuge for a large entomofauna. In addition to the bruchid beetle of *Acacia sp.* (*P. spinipes)*, partial dehiscence may be a favorable condition for attack by a second bruchid (*S. furcatus*), which normally attacks seeds fallen on the substrate. This bruchid community may provide a reserve of parasitoids which could be used for biological control of Bruchidae.

**Acknowledgements**

The pods were collected during two missions, the first as part of an FAO study into certain aspects of biological control of *Bruchus pisorum* (Chile in 1999) , and the second as part of a lecture tour at the Faculty of Agronomy of Montevideo (funded by the Cultural Department of the French Embassy and the University of the Republic of Uruguay). I would like to thank those in charge of the programs in Chile (Ing. Dr Marcos Gerding) and in Uruguay (Ing. Dr Basso César), who helped me to conduct research on the parasitoid communities of the bruchid beetles in South America.   Thanks also to Dr Dan Johnson, Gérard Delvare, John Lasalle, Jean-Luc Mercier and J. Yves Rasplus for their help in determining the different insect species.

The English text was corrected by Elizabeth Yates (Inter-connect).]

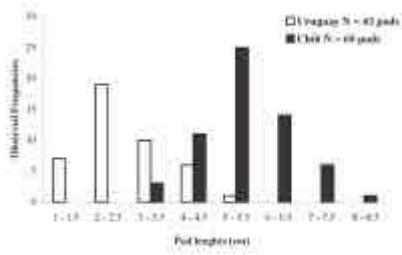

Figure 1. Pod lengths of persistent *A. caven* collected during the spring in Chili and Uruguay.

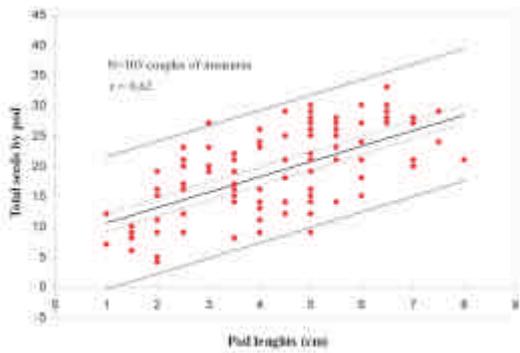

Figure 2. Relationship between the pod length and the number of seeds per pod.

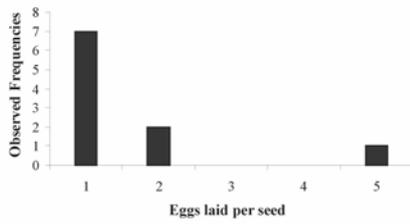

Figure 3. Egg-laying of *Stator furcatus* on seeds of *A. caven*. After a rainy spring the partially dehiscent pods permitted the egg-laying on seeds by *S. furcatus*.

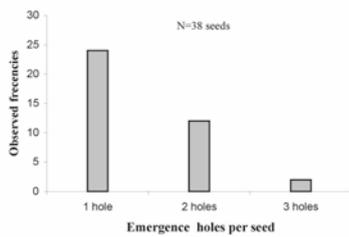

Figure 4. Emergence holes pierced per seed by *Stator furcatus*. From one to three *S. furcatus* could develop per seed.

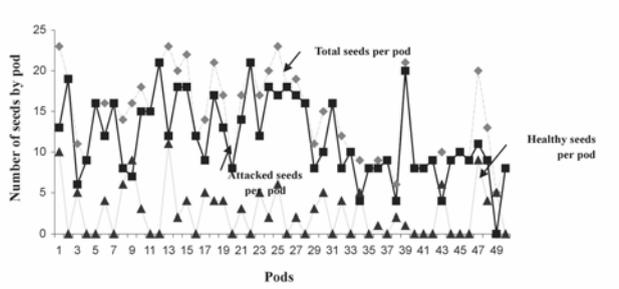

**Figure 5.** Evolution of seed quality per pod of *A.caven*. The opening of 50 *A.caven* pods permit to separate attacked seeds by the Bruchidae and to evaluate the healthy seeds able to start to growth.

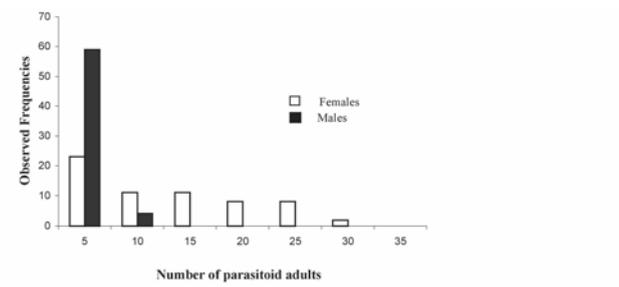

**Figure 6.** Total offspring of 63 egg-laying *Monoska dorsiplana* inseminated females. Each egg-laying female was provided with 1 host per day for 4 consecutive days.

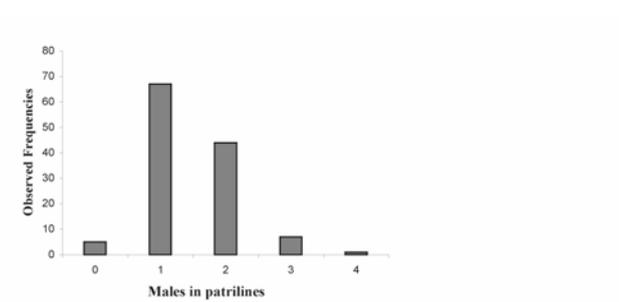

**Figure 7.** Males distribution of *Monoksa dorsiplana* in the patrilines of one egg-laying female. The most common patriline was composed by 1 male.

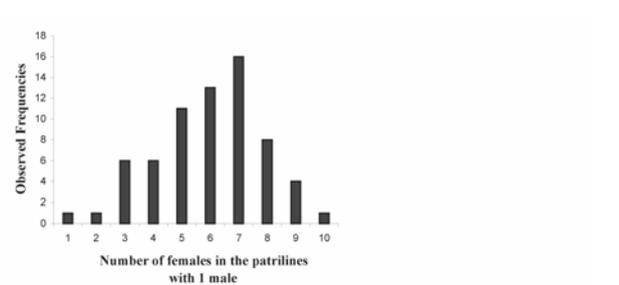

**Figure 8.** Females emerged per host parasitized by one egg-laying *Monoksa dorsiplana* female